\documentstyle[12pt,aps,epsf]{revtex}
\draft
\def\beginwide{
        \end{multicols} \vspace*{-0.5cm} \noindent
        \rule{3.5in}{.1mm}\rule{.1mm}{5mm} \widetext \medskip }
\def\beginwidetop{
        \end{multicols} \vspace*{-0.5cm} \noindent
        \widetext \medskip }
\def\endwide{
        \hspace*{3.35in}~\rule[-5mm]{.1mm}{5mm}\rule{3.5in}{.1mm}
        \begin{multicols}{2} \vspace*{-1.0cm} \noindent }
\def\endwidebottom{
        \begin{multicols}{2} \vspace*{-1.0cm} \noindent }

\newcommand{\beq}{\begin{equation}}
\newcommand{\eeq}{\end{equation}}
\newcommand{\bdis}{\begin{displaymath}}
\newcommand{\edis}{\end{displaymath}}
\newcommand{\bea}{\begin{eqnarray}}
\newcommand{\eea}{\end{eqnarray}}
\newcommand{\barr}{\begin{array}}
\newcommand{\earr}{\end{array}}
\begin{document}

\title{Renormalization of 
Systems with Non-equilibrium Critical Stationary States}

\author{Alessandro Vespignani$^{(1)}$, Stefano Zapperi$^{(2)}$ 
and Vittorio Loreto$^{(3)}$}

\address{1) Instituut-Lorentz, University of Leiden, P.O. Box 9506
2300 RA, Leiden, The Netherlands\\
2)Center for Polymer Studies and Department of Physics
Boston University, Boston, MA 02215 \\
3)ENEA Research Center, loc. Granatello, C.P.32, 
80055 Portici, Napoli, Italy}

\date{\today}

\maketitle
\begin{abstract}
We introduce the general formulation of a renormalization 
method suitable to study the  critical properties of 
non-equilibrium systems with steady-states: the Dynamically
Driven Renormalization Group. 
We renormalize the time evolution operator by computing
the rescaled time transition rate between coarse grained states.
The obtained renormalization equations are coupled to a stationarity
condition which provides the approximate non-equilibrium statistical weights 
of steady-state configurations to be used in the calculations. 
In this way we are able to write
recursion relations for the parameters evolution
under scale change,
from which we can extract numerical values for the critical exponents.
This general framework  allows the systematic analysis of 
several models showing self-organized criticality in terms of usual concepts
of phase transitions and critical phenomena.
\end{abstract}


%
%
%

In the last decade non-equilibrium critical phenomena have 
attracted a wide interest in statistical physics. 
Critical systems are characterized by the absence of a 
characteristic lengthscale, strong fluctuations and 
non-analyticity of the correlation functions. 
Examples of this behavior can be found in 
phase transitions \cite{domb,katz,zia}, 
self-organized critical (SOC) systems \cite{soc}, 
fractal growth \cite{vic} and a vast class of 
complex systems\cite{bb}. 
The major source of difficulties in the study of
non-equilibrium critical phenomena\cite{zia,rev}
lies in the absence of a general criterion,
like the use of the Gibbs distribution in equilibrium systems,
to assign an ensemble statistical measure to a particular 
configuration of the system.
The probability distribution is instead a time dependent solution
of a master equation, which only in some particular cases 
becomes stationary in the long time limit.

In this letter we present the general formalism 
of a  real space dynamical renormalization 
group (RG) scheme for systems with a non-equilibrium critical steady-state: 
the Dynamically Driven Renormalization Group
(DDRG). 
The method combines the renormalization of the time evolution operator 
with a stationarity condition which allows the calculation 
of the approximate steady-state configurations
probability distribution. This coupling acts at each coarse 
graining step and therefore represents a {\em driving} for 
the renormalization group equations. 
For SOC systems \cite{bak2,bak,ds}, the DDRG allows us to derive 
in a broader framework previous RG schemes \cite{pvz,lpvz,iva} 
and to formulate a more systematic approach.
Here we show the explicit application of the 
DDRG to the FFM \cite{bak,ds}, which we can now study in the whole 
parameters space.
Possible applications of the DDRG are not restricted to SOC models: 
the method can be used to study other equilibrium or  non-equilibrium critical
phenomena such as driven diffusive systems \cite{katz,zia}, 
which to our knowledge have never been 
approached by real space RG methods.

We consider discrete lattice models on a 
$d$-dimensional lattice. To each site $i$
is associated a variable $\sigma_i$, which can assume $q$ different values
($\sigma_i= 0,1,\cdots,q$). A complete set $\sigma\equiv\{\sigma_i\}$ of 
lattice variables specifies a configuration of the system. 
We define 
$\langle\sigma\mid T(\mu) \mid \sigma^0\rangle$
as the transition rate
from a configuration $\sigma_0$ to
a configuration $\sigma$ in a time step $t$ 
as a function of a set of parameters  $\mu=\{\mu_i\}$.
The time dependent
probability distribution $P(\sigma,t)$ for the configurations 
of the system, obeys the following master equation (ME)
\beq
P(\sigma,t_0+t)=\sum_{\{\sigma^0\}} \langle\sigma\mid 
T(\mu) \mid \sigma^0\rangle
P(\sigma^0,t_0).
\eeq
The explicit solution of the master equation is in general
not available but we can extract the critical properties
of the model by a renormalization group analysis.
We coarse grain the system by rescaling lengths and time according to  
the transformation $x\to bx$ and $t\to b^z t$. 
The renormalization transformation is 
constructed through an operator  
${\cal R} (S,\sigma)$ that 
introduces a set of coarse grained variables $S\equiv\{S_i\}$ 
and rescales the lengths of the system \cite{nv}. 
In general, $\cal R$ is a projection operator with the properties
${\cal R}(S,\sigma)\geq 0$ for any $\{S_i\},\{\sigma_i\}$, and 
$\sum_{\{S\}} {\cal R}(S,\sigma)=1$.
These properties preserve the normalization condition
of the renormalized distribution. The explicit form of the operator $\cal R$
is defined case by case in the various applications of the method. 
Usually, it corresponds to a block transformation in which lattice 
sites are grouped together in a super-site 
that  defines the renormalized 
variables $S_i$ by means of a majority or spanning rule.

We subdivide the time step in intervals of the unitary time scale ($t_0=0$) 
obtaining the coarse graining of the system as follows:
\beq
P^{\prime}(S,t')= 
\sum_{\{\sigma\}}{\cal R}(S,\sigma)
\sum_{\{\sigma^0\}}\langle\sigma\mid T^{b^z}(\mu)\mid \sigma^0\rangle
P(\sigma^0,0)
\label{start}
\eeq
where we have included the application of the operator 
$\cal R$ and $t'=b^z t$. The meaning
of $\langle\sigma\mid T^{b^z}(\mu) \mid \sigma^0\rangle$
has to be defined explicitly:
the simplest possibility is $b^z=N$ where  $N$ is an integer number, 
and $T^N$ denotes the application of the dynamical operator $N$ times.
In general, since we are dealing with a discrete time 
evolution, we have to consider $T^{b^z}$ as a convolution over
different paths, chosen by an appropriate condition. 
The detailed definition of the effective operator $T^{b^z}$
is reported in Ref.~\cite{ddrg}. 
By multiplying and dividing each term of eq.~(\ref{start})
by $P^{\prime}(S^0,0)=\sum_{\{\sigma^0\}}{\cal R}(S^0,\sigma^0)
P(\sigma^0,0)$ and using the properties of the operator ${\cal R}$, 
we get after some algebra:
\beq
P^{\prime}(S,t')=
\sum_{\{S^0\}}(\frac
{\sum_{\{\sigma^0\}}\sum_{\{\sigma\}}
{\cal R}(S^0,\sigma^0)
{\cal R}(S,\sigma)\langle\sigma\mid T^{b^z}(\mu) \mid \sigma^0\rangle 
P(\sigma^0,0)}
{\sum_{\{\sigma^0\}}{\cal R}(S^0,\sigma^0)
P(\sigma^0,0)})P^{\prime}(S^0,0)
\label{mefine}
\eeq
which finally identifies  the renormalized dynamical operator
$\langle S\mid T'\mid S^0\rangle$. In other words the new dynamical 
operator $T'$ is the sum over all the dynamical paths of $b^z$ steps
that from a starting configuration $\{\sigma^0_i\}$ lead to a configuration 
$\{\sigma_i\}$, which renormalize respectively in $\{S^0_i\}$ and $\{S_i\}$.
The sum is weighted by the normalized 
statistical distribution of each configuration.

We apply this scheme to systems with a steady-state described by 
a stationary distribution $P(\sigma,t\to\infty)=W(\sigma)$.
For equilibrium systems the stationary distribution has the
Gibbs form $W(\sigma)\sim \exp(-\beta H(\sigma))$, where
$H(\sigma)$ is the Hamiltonian.
There is not such a general criterion for non-equilibrium 
dynamical system, therefore
we have developed an approximate method to evaluate 
the stationary distribution to be used in the calculation of 
the renormalized master equation. 
The simplest approximation 
considers only the incoherent part of the stationary distribution 
which does not include correlations and can therefore be factorized.
For systems characterized by a q-state variables it has the form 
\beq
W^{\mbox{(i)}}(\sigma)= \prod_{i}\langle\rho_{\sigma_i}\rangle
\label{produ}
\eeq
where $\langle\rho_\kappa\rangle$ is the average density of sites in the 
$\kappa$-state. In this way, we have approximated the
probability of each configuration 
$\{\sigma_i\}$ as the product measure of the mean field  probability to 
have a state $\sigma_i$ in each corresponding site.
The values of the densities $\{\langle\rho_\kappa\rangle\}$
as a function of the parameters $\mu$
are obtained by solving appropriate mean-field equations
in the long time limit. These equations have the form of
a stationarity condition
\beq
\frac{\partial}{\partial t}\{\langle\rho_\kappa\rangle\}=
{\cal S}_\mu (\{\langle\rho_\kappa\rangle\})=0
\label{stat}
\eeq
where the operator ${\cal S}_\mu$  describes the evolution
of the system as a function of the dynamical parameters defined
above. Time independent solutions of Eq.~(\ref{stat}) will be referred to as 
``steady-states'', although we should keep in mind that those are only the 
average states of the ensemble \cite{kei}. 
In ordinary statistical systems, Eq.~(\ref{stat}) represents the 
thermodynamic equilibrium condition. For driven dynamical systems, 
it describes the {\em driving} of
the system to the non-equilibrium steady-state, 
by means of a balance condition.

By inserting this approximate distribution 
in Eq.~(\ref{mefine}), we obtain the renormalized dynamical operator
\beq
\langle S\mid T'(\mu) \mid S^0\rangle=
\frac{\sum_{\{\sigma^0\}}\sum_{\{\sigma\}}
{\cal R}(S^0,\sigma^0)
{\cal R}(S,\sigma)\langle\sigma\mid T^{b^z}(\mu) \mid \sigma^0\rangle 
\prod_{i}\langle\rho_{\sigma_i^0}\rangle}
{\sum_{\{\sigma^0\}}{\cal R}(S^0,\sigma^0)
\prod_{i}\langle\rho_{\sigma_i^0}\rangle}
\label{ddrg1}
\eeq
where the densities are calculated at each coarse graining step
from the stationary condition (Eq.\ref{stat}) with the corresponding 
renormalized dynamical parameters $\{\mu\}$.
Since in this framework Eq.(\ref{stat}) drives the RG equations 
acting as a feedback on the scale transformation, we call it the 
{\em driving  condition}.

The Eq.s~(\ref{ddrg1}),(\ref{stat}) are the basic renormalization equation
from which the desired recursion relations are obtained. 
Imposing that the renormalized
operator $T'$ has the same functional form of the operator $T$, 
i.e. $T'(\mu)=T(\mu')$, we obtain the rescaled 
parameter set $\mu'=f(\mu)$. This implies that the renormalized  single time 
distribution $P^{\prime}(S,t')$ has the same functional form of the original 
distribution $P(\sigma,t)$. The critical behavior of the model 
is obtained by studying the fixed points $\mu^*=f(\mu^*)$.
Since  we are dealing with discrete evolution operators $T$, we define 
the time scaling factor $b^z$ as the average number of steps we  
apply the operator $T$ in order to obtain that $T'(\mu)=T(\mu')$ for the 
coarse grained system. 
In this way we obtain a time recursion 
relation $t'=g(\mu)t$, or equivalently $b^z=g(\mu)$,
 from which it is possible 
to calculate the dynamical critical exponent 
$z={\log g(\mu^*)}/{\log b}$.
In this form of the DDRG, we take into account only the uncorrelated part 
of the steady-state probability distribution. The results obtained 
are not trivial because correlations in the systems are considered in 
the dynamical renormalization of the operator $T$, that given a starting 
configuration traces all the possible paths leading to the renormalized 
final configuration. Moreover, geometrical correlations are 
treated by the operator $\cal R$ that maps the system by means of 
spanning conditions or majority rules. 
The renormalized uncorrelated part of the stationary distribution is  
evaluated from the stationary condition with renormalized parameters,
thus providing an effective treatment of correlations.
One can then improve the results by including higher order
contributions to the unknown stationary distribution $W(\sigma)$
using cluster variation methods \cite{dick}. Naturally the above scheme 
can also be applied to equilibrium critical pheneomena, where the driving
condition is represented by the equilibrium mean field equations \cite{ddrg}.

The DDRG is a useful tool to study the critical properties of
SOC systems. In fact, these systems evolve 
spontaneously in a scale invariant stationary state. 
The Forest Fire Model is a simple automaton which has 
been introduced by Bak et al. \cite{bak} as an example of SOC, and has been 
then modified by Drossel and Schwabl \cite{ds}.
The model is defined on a lattice in which
each site can be empty ($\sigma_i=0$), occupied by a green tree 
($\sigma_i=1$) 
or by a burning tree ($\sigma_i=2$). At each time step the 
lattice is updated as follows: i) a burning tree becomes an empty site;
ii) a green tree becomes a burning tree if at least one of its 
neighbors is burning; iii) a tree can grow in an empty site 
with probability $p$; iv) a tree without burning nearest neighbors 
becomes a burning tree with probability $f$.
The model was first studied in the case $f=0$ for the limit of very 
slow tree growth ($p\to 0$). In this limit the 
critical behavior is trivial: the model shows spiral-shaped fire fronts
separated by a diverging length $\xi\sim p^{-\nu_p}$, where
$\nu_p\simeq 1$ \cite{gras}.
In the case $f>0$, the system is supposed to exhibit SOC
under the hypothesis of a double separation of time scales:
trees grow fast compared with the occurrence of lightnings  
and forest clusters burn down much faster than trees grow. 
This request is expressed by the double limit 
$\theta\equiv f/p \rightarrow 0$ and  $p \rightarrow 0$.
The critical state is characterized by a power law distribution 
$P(s)=s^{-\tau}$ of the forest clusters of $s$ sites 
(avalanches in the SOC terminology) and  the average  
cluster radius (the correlation length) scales as $R \sim \theta^{-\nu_R}$. 

With the DDRG framework we are able to generalize a previous RG scheme
\cite{lpvz}, in order to include the proper treatment of the 
time scaling and to study the limit $f=0$ (deterministic FFM).
The dynamical rules of the FFM are local and the set of dynamical 
parameters, defined by $\mu=\{f,p\}$, 
is obtained explicitly in terms of the dynamical operators
acting on a single site, i.e. $\langle 1\mid T\mid 0\rangle = p$ and 
$\langle 2\mid T\mid 1\rangle = f$. The  
relevant dynamical scales is defined by the burning process which occurs 
with probability one. 
We define a cell-to-site transformation 
with scale factor $b=2$ or larger. 
The rules defining the cell renormalization operator ${\cal R}$ are 
standard geometrical spanning conditions \cite{cres}, and their explicit form
can be found in Ref.\cite{ddrg}.
The above scheme defines a finite lattice truncation on four (two) 
sites cells in $d=2$ ($d=1$), and denoting by an index $\alpha$ each
cell configuration, we have that $\sum_{\{\sigma_i\}}\to 
\sum_{\alpha}$. The renormalization equations that define 
the renormalized parameters can be conveniently written
as 
\beq
\langle S_i\mid T'\mid S_i^0\rangle= \frac{\sum_{\alpha}
\sum_{\alpha'}
\langle\alpha'\mid T^{b^z}\mid \alpha\rangle W_{\alpha}}
{\sum_{\alpha} W_{\alpha}}
\eeq
where $\mid\alpha\rangle$ and $\mid\alpha'\rangle$ 
are the cell states which renormalize respectively 
in $\mid S_i^0\rangle$ and  $\mid S_i\rangle$. We keep the subscript $i$
since the states refer now to a single coarse
grained site and not to a configuration of the system.
With $W_\alpha$ we denote the stationary
statistical weight of each $\alpha$ configuration. 
This distribution is approximate following the DDRG scheme in the lowest 
order (Eq.4), in which the average steady-state densities 
$\langle\rho_\kappa\rangle$
are obtained as a function of $\mu=\{f,p\}$ from the stationary 
solution of dynamical  mean field equations \cite{cris}.

We focus our analysis in the critical region  denoted by the condition 
$f\ll p\ll 1$, namely where the system shows critical behavior.
The time scaling factor is obtained by imposing that the
renormalized burning process occurs with probability one
($\langle 0\mid T'\mid 2\rangle = 1$).
In $d=1$ this condition is fulfilled up to second order
in $f$ and $p$ and gives $z=1$,
recovering the exact result of Ref. \cite{exact}.
This result is due to the fact that in $d=1$ there is
only a possible way to span the cell, and consequently no 
proliferations are generated.
In $d=2$ one has to consider the
average over different paths, and new dynamical interactions are generated 
at each RG step. This is a signature that we need an approximation which
truncates the parameter space after each iteration so that it remains closed.
This is done by considering just the leading order in $f$ and $p$ in the 
renormalization equations, and ignoring any proliferations generated at each
group iteration.
With this scheme we obtain $z=1$, which is not an exact result also if in 
good agreement with numerical simulations ($z=1.04$ \cite{clar}).
It is worth to remark that the DDRG allows to overcome the approximations 
present in the approach of Ref.\cite{lpvz}, where the time scaling was 
not properly considered because of the assumption of
an infinite time scale separation. In addition the general scheme shown so
far provides the inroad towards a systematic improvement of the results
by introducing higer order correlations in the stationary distribution as 
discussed in Ref.\cite{patz}.  

Once the time scale factor
is set we can write recursion relations for $p$ and $f$, or 
equivalently $\theta'=x(\theta,p)$ 
and $p'=y(\theta,p)$,
evaluating the probabilities that a coarse grained
cell grows or is struck by a lightning in $b^z$ steps.
The  driving condition and recursion relations 
derivation is  long and tedious and the explicit equations are reported 
elsewhere \cite{ddrg}. The flow diagram  
is stable with respect to different coarse graining rules,
and for $d=1$ and $d=2$ we find
a repulsive fixed point in $\theta_c=0$ and $p_c=0$.
The fixed point densities 
are obtained from the driving condition and depend on the 
dimensionality. In order to discuss the critical behavior we have to 
linearize the recursion relations in the proximity of this fixed point and to 
find the relevant eigenvalues of the diagonal transformation:
\beq
\lambda_1=\left. \frac{\partial\theta'}{\partial\theta}\right|_{\theta_c,p_c}
\mbox{\hspace{5mm}}; \mbox{\hspace{5mm}}
\lambda_2=\left. \frac{\partial p'}{\partial p}\right|_{\theta_c,p_c}
\eeq
In $d=2$ the largest eigenvalue is given by 
$\lambda_1$, which 
determines the leading scaling exponent $\nu_R=\log b/\log \lambda_1=0.7$
(for $b=2$) obtained in Ref.\cite{lpvz}. The result is in good agreement
with numerical simulation ($\nu_R=0.6$ \cite{clar}.
In the limit $f=0$ the critical behavior is 
governed by the second eigenvalue $\lambda_2$. This eigenvalue 
and its relative exponent describes  the behavior of the correlation 
length in  the deterministic FFM. As opposed to $\lambda_1$,
the value of $\lambda_2$ depends on the absolute
value of the time scaling factor \cite{commdros}, and therefore could not
be obtained without the DDRG formalism.
The numerical value we obtain in $d=1,2$ is 
$\nu_p=\log 2/\log \lambda_2=1.0$, which is in 
excellent agreement with 
the simulation results $\nu_p\simeq 1$ \cite{gras}. 

Our characterization of the flow diagram clarifies the 
critical nature of the model. The FFM  is 
critical only for $\theta_c=0,p_c=0$. This implies
that $\theta,p$ are the {\em control parameters} of the model,  
and the critical state is reached only by a fine tuning 
of these parameters. Similar results are obtained by applying the DDRG 
to the sandpile model \cite{ddrg}. These results allows us
to clarify the meaning of SOC with respect to
non equilibrium critical phenomena.
In SOC literature it is often reported that the 
origin of scale invariance in nature lies 
in the absence of tuning parameter, like 
the  critical temperature in Ising models. 
In the renormalization group language this would 
imply that no relevant parameters should
be present. The situation is, however, more subtle.
It has been recognized that a common characteristic of SOC
systems is the presence of two time scales 
$\tau_a$, the typical relaxation (activity) time, and $\tau_d$ 
the external driving time scale (often an external noise).
In order to observe criticality the ratio ${\cal T}=\tau_a/\tau_d$ 
must be vanishingly small (${\cal T}\to 0$) \cite{commdros,grin}. 
With our approach we can recast the above concept in more  
formal terms. Our RG analysis shows the time scales ratio 
${\cal T}$ is indeed the {\em control parameter} of SOC models. 
This parameter is the ratio  between $f$, $p$ and the burning time scale
in the Forest Fire model or the sand addition and 
the avalanche dissipation in sandpiles,
but is always related to the ratio between different time scales.
{}From a theoretical point of view the critical nature of 
SOC systems is not different from that of non-equilibrium phase transitions.
The peculiarity of these systems is that close to the critical point 
the system is quite stable to changes of the dynamical time scales. 
In fact, the reduced control parameter which is defined as 
$\epsilon=({\cal T}-{\cal T}_c)/{\cal T}_c$, in SOC system is ${\cal T}$
itself, being ${\cal T}_c=0$. 
This implies that if $\epsilon\simeq 0$, 
even relevant changes of the control parameter
(${\cal T}\to n{\cal T}$ and $n < \epsilon^{-1}$) do not drive the 
system far from the critical region.
Apparently the system would not be affected  by changes 
of ${\cal T}$, and in  
this sense SOC systems are not very sensitive 
to fine tuning of the control 
parameter. The meaning of SOC is then related to the widespread 
existence of phenomena ruled by very different time scales and not to
the absence of relevant control parameters as often reported in literature.

A.V. is indebted with J.M.J. van Leeuwen for very interesting discussions.
The Center for Polymer Studies is supported by NSF.

\end{document}